# On the Casimir-Lifshitz Force in a System of Two Parallel Plates in Relative Nonrelativistic Motion

A.A. Kyasov, G.V. Dedkov

Nanoscale Physics Group, Kabardino-Balkarian State University, Nalchik, Russia

In this paper we correct a previously obtained expression for the Casimir-Lifshitz attraction force (Dedkov, Kyasov, Surf. Sci., 2010) in a system of two parallel plates in relative nonrelativistic motion, assuming the total thermal equilibrium of the system.

In our papers [1] we have established the relationships between the two basic geometrical configurations when calculating fluctuation-electromagnetic interactions, namely the configuration of a small neutral dipole particle above a flat surface (configuration "1") and a configuration of two parallel plates (configuration "2"), assuming that the whole system is in total thermal equilibrium with temperature $T$. A particle and one of the plates in configuration 2 are assumed to be in relative nonrelativistic motion with respect to the resting plate (see Fig. 1).

Our analysis was based on the correspondence rule between configurations 1 and 2 which allows one to calculate the fluctuation-electromagnetic forces and the rate of radiative heat exchange in both configurations using transitions $1 \to 2$ or $2 \to 1$. Transition $2 \to 1$ is known since the pioneering work by Lifshitz [2] when it has been applied for calculating the Casimir-Polder force in static configuration 1. This transition is implemented using the limiting relation of rarified medium $\varepsilon_1(\omega) - 1 = 4\pi n_1 \alpha_1(\omega) \to 0$, where $\varepsilon_1(\omega)$ and $n_1$ are the dielectric permittivity and atomic density of the material of the first (for definiteness) plate (see Fig. 1b). According to [3], the following relation takes place

$$F_z^{(1)}(z) = \frac{1}{n_1 S} \frac{dF_z^{(2)}(l)}{dl}\bigg|_{l=z} \qquad (1)$$

where $F_z^{(2)}(l)/S$ describes the Casimir-Lifshitz force per unit area of two parallel plates with surface area $S$, separated by a gap of width $l$, and $F_z^{(1)}(z)$ is the Casimir-Polder force between a polarizable particle and the resting plate.

In [1], we applied Eq. (1) to transition $2 \to 1$ in a situation out of dynamical equilibrium for obtaining the dynamical Casimir-Lifshitz force $F_z^{(2)}(l)$ in the case when the upper plate moves with nonrelativistic velocity $V \ll c$. Apart from (1), in formulating the proper correspondence rule we used the exact expression for $F_z^{(1)}(z)$ at $V \neq 0$ [1] and the exact expression for Casimir-Lifshitz force $F_z^{(2)}(l)$ at $V = 0$ [2]. The latter is given by (omitting superscript "(2)" for brevity) [3]

$$F_z(l) = -\frac{\hbar S}{4\pi^3} \mathrm{Im} \int_0^\infty d\omega \int_{-\infty}^{+\infty} dk_x \int_{-\infty}^{+\infty} dk_y \, q_0 \cdot \coth\left(\frac{\hbar \omega}{2k_B T}\right) \cdot$$
$$\cdot \left[ \left(\frac{1}{\Delta_{1e}} \frac{1}{\Delta_{2e}} \exp(2q_0 l) - 1\right)^{-1} + \left(\frac{1}{\Delta_{1m}} \frac{1}{\Delta_{2m}} \exp(2q_0 l) - 1\right)^{-1} \right], \quad (2)$$

$$\Delta_{ie} = \frac{q_0 \varepsilon_i(\omega) - q_i}{q_0 \varepsilon_i(\omega) + q_i}, \quad \Delta_{im} = \frac{q_0 \mu_i(\omega) - q_i}{q_0 \mu_i(\omega) + q_i}, \quad q_i = \left(k^2 - \frac{\omega^2}{c^2} \varepsilon_i(\omega)\mu_i(\omega)\right)^{1/2},$$
$$q_0 = \left(k^2 - \frac{\omega^2}{c^2}\right)^{1/2}, \quad k^2 = k_x^2 + k_y^2 \quad (3)$$

where $i = 1, 2$, $\varepsilon_i(\omega)$ and $\mu_i(\omega)$ are the dielectric permittivity and magnetic permeability of the plate materials (upper plate 1 and lower plate 2).

Using the identity

$$\left(\frac{1}{\Delta_1} \frac{1}{\Delta_2} \exp(2q_0 l) - 1\right)^{-1} = \frac{\Delta_1 \Delta_2 \exp(-2q_0 l) - |\Delta_1|^2 |\Delta_2|^2 \exp[-2(q_0 + q_0^*)l]}{|1 - \exp(-2q_0 l)\Delta_1 \Delta_2|^2} \quad (4)$$

Eq. (2) takes the form which is convenient for other transformations

$$F_z(l) = -\frac{\hbar S}{4\pi^3} \int_0^\infty d\omega \int_{-\infty}^{+\infty} dk_x \int_{-\infty}^{+\infty} dk_y \, \coth\left(\frac{\hbar \omega}{2k_B T}\right) \cdot$$
$$\cdot \sum_{j=e,m} \left\{ \frac{\mathrm{Im}\,\Delta_{1j} \cdot \mathrm{Re}[q_0 \exp(-2q_0 l)\Delta_{2j}] + \mathrm{Re}\,\Delta_{1j} \cdot \mathrm{Im}[q_0 \exp(-2q_0 l)\Delta_{2j}]}{|1 - \exp(-2q_0 l)\Delta_{1j}\Delta_{2j}|^2} - \frac{|\Delta_{1j}|^2 |\Delta_{2j}|^2 \,\mathrm{Im}[q_0 \exp(-2(q_0 + q_0^*)l)]}{|1 - \exp(-2q_0 l)\Delta_{1j}\Delta_{2j}|^2} \right\} \quad (5)$$

The second term in figure brackets of (5) differs from zero only at $k < \omega/c$ and therefore in the nonrelativistic approximation ($c \to \infty$) it is absent. Moreover, this term disappears upon the limiting transition to a rarified medium for one of the plates, and due to this it was omitted in [1]. By performing transition to the moving first plate in (5) in full agreement with [1] one obtains

$$F_z(l,V) = -\frac{\hbar S}{4\pi^3} \int_0^\infty d\omega \int_{-\infty}^{+\infty} dk_x \int_{-\infty}^{+\infty} dk_y \cdot$$

$$\cdot \sum_{j=e,m} \left\{ \frac{\mathrm{Im}\Delta_{1j}(\omega^+) \cdot \mathrm{Re}[q_0 \exp(-2q_0 l)\Delta_{2j}(\omega)]}{|1-\exp(-2q_0 l)\Delta_{1j}(\omega^+)\Delta_{2j}(\omega)|^2} \coth\left(\frac{\hbar\omega^+}{2k_B T}\right) + \right.$$
$$\left. + \frac{\mathrm{Re}\Delta_{1j}(\omega^+) \cdot \mathrm{Im}[q_0 \exp(-2q_0 l)\Delta_{2j}(\omega)]}{|1-\exp(-2q_0 l)\Delta_{1j}(\omega^+)\Delta_{2j}(\omega)|^2} \coth\left(\frac{\hbar\omega}{2k_B T}\right) \right\} - \qquad (6)$$

$$-\frac{\hbar S}{4\pi^3} \int_0^\infty d\omega \int_{k<\omega/c} d^2k |q_0| \cdot \sum_{j=e,m} \frac{|\Delta_{1j}(\omega^+)|^2 |\Delta_{2j}(\omega)|^2}{|1-\exp(-2q_0 l)\Delta_{1j}(\omega^+)\Delta_{2j}(\omega)|^2} \coth\left(\frac{\hbar\omega}{2k_B T}\right)$$

where $\omega^+ = \omega + k_x V$ and $V \ll c$. It is worth noting that expressions for tangential force $F_x(l,V)$ and heating rate $\dot{Q}(l,V)$ obtained in [1] remain the same. In order to obtain the expression for $F_z^{(1)}(z,V)$ in configuration 1, one should substitute (6) in (1), performing the transitions [1]

$$\Delta_{1e}(\omega) \to \frac{\pi n_1}{q_0^2} \left[\alpha_e(\omega)(2k^2 - \omega^2/c^2) + \alpha_m(\omega)\omega^2/c^2\right]$$

$$\Delta_{1m}(\omega) \to \frac{\pi n_1}{q_0^2} \left[\alpha_m(\omega)(2k^2 - \omega^2/c^2) + \alpha_e(\omega)\omega^2/c^2\right]$$

These relations should be used instead $\varepsilon_1(\omega) - 1 = 4\pi n_1 \alpha_e(\omega) \to 0$ and $\mu_1(\omega) - 1 = 4\pi n_1 \alpha_m(\omega) \to 0$ in static case $V = 0$. A remarkable feature of the second integral term in (6) is its minor dependence on the distance $l$.

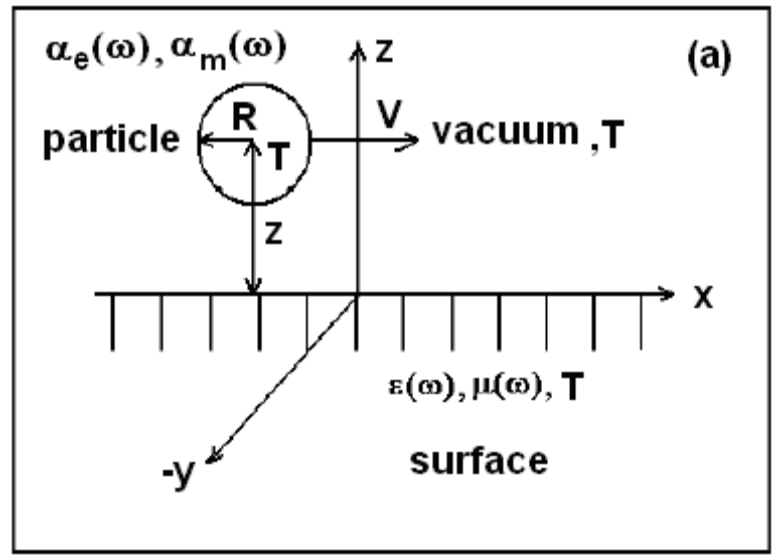

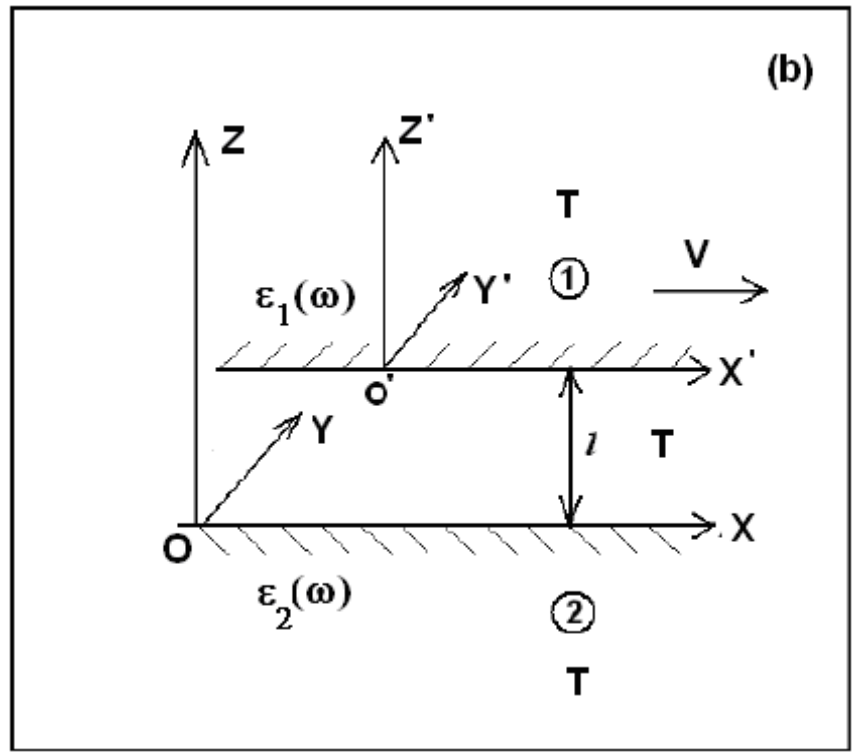

Fig. 1. Configuration 1 (a) and configuration 2 (b).